\documentclass[num-refs,11pt,table,dvipsnames]{wiley-article}

\usepackage{xcolor}
\usepackage{listings}
\usepackage[frozencache=true,cachedir=.]{minted}
\usepackage{newunicodechar}

\newunicodechar{ρ}{{$\rho$}}
\newunicodechar{λ}{{$\lambda$}}

\usepackage{siunitx}

\usepackage{tikz}
\usepackage{pgfplots}
\usepackage{pgfplotstable}
\usetikzlibrary{shapes,arrows,calc,fit,backgrounds}
\usetikzlibrary{positioning}
\usetikzlibrary{arrows,arrows.meta}
\usepgflibrary{shapes.multipart}

\lstdefinelanguage{julia}
{
  keywordsprefix=\@,
  morekeywords={
    exit,whos,edit,load,is,isa,isequal,typeof,tuple,ntuple,uid,hash,finalizer,convert,promote,
    subtype,typemin,typemax,realmin,realmax,sizeof,eps,promote_type,method_exists,applicable,
    invoke,dlopen,dlsym,system,error,throw,assert,new,Inf,Nan,pi,im,begin,while,for,in,return,
    break,continue,macro,quote,let,if,elseif,else,try,catch,end,bitstype,ccall,do,using,module,
    import,export,importall,baremodule,immutable,local,global,const,Bool,Int,Int8,Int16,Int32,
    Int64,Uint,Uint8,Uint16,Uint32,Uint64,Float32,Float64,Complex64,Complex128,Any,Nothing,None,
    function,type,typealias,abstract
  },
  sensitive=true,
  morecomment=[l]{\#},
  morestring=[b]',
  morestring=[b]" 
}

\papertype{Original Article}
\paperfield{Journal Section}

\definecolor{ibilight}{RGB}{193,216,237}
\definecolor{ibidark}{RGB}{0,73,146}
\definecolor{uke2}{RGB}{170,156,143}
\definecolor{uke3}{RGB}{87,87,86}
\definecolor{ukesec1}{RGB}{255,223,0}
\definecolor{ukesec2}{RGB}{239,123,5}
\definecolor{ukesec3}{RGB}{104,195,205}
\definecolor{ukesec4}{RGB}{138,189,36}
\definecolor{tuhh}{RGB}{45,198,214}

\title{MRIReco.jl: An MRI Reconstruction Framework written in Julia}

\author[1,2\authfn{1}]{Tobias Knopp}
\author[1,2\authfn{1}]{Mirco Grosser}

\corremail{tobias.knopp@tuhh.de}
\corraddress{Tobias Knopp, Lottestrasse 55, 22529 Hamburg, Germany}

\affil[1]{Institute for Biomedical Imaging, Hamburg University of Technology, Hamburg, Germany}
\affil[2]{Section for Biomedical Imaging, University Medical Center Hamburg-Eppendorf, Hamburg, Germany}

\fundinginfo{-}

\wordcount{5996}

\runningauthor{T. Knopp et al.}

\newcommand{\zb}[1]{\mbox{\boldmath{${#1}$}}}
\newcommand{\zbs}[1]{\mbox{\boldmath\scriptsize{${#1}$}}}

\renewcommand{\d}{\, \text{d}}

\def\i{\mbox{\rm i}}
\def\ti{\mbox{\scriptsize{\rm i}}}

\begin{document}
\fontsize{9}{11.5}\selectfont

\maketitle

\begin{abstract}
\textbf{Purpose:}
The aim of this work is to develop a high-per\-for\-man\-ce, flexible and easy-to-use MRI reconstruction framework using the scientific programming language Julia.\\
\textbf{Methods:}\\
Julia is a modern, general purpose programming language with strong features in the area of signal / image processing and numerical computing. It has a high-level syntax but still generates efficient machine code that is usually as fast as comparable C/C++ applications. In addition to the language features itself, Julia has a sophisticated package management system that makes proper modularization of functionality across different packages feasible. Our developed MRI reconstruction framework MRIReco.jl can therefore reuse existing functionality from other Julia packages and concentrate on the MRI-related parts. This includes common imaging operators and support for MRI raw data formats. \\
\textbf{Results:}\\
MRIReco.jl is a simple to use framework with a high degree of accessibility. While providing a simple-to-use interface, many of its components can easily be extended and customized.
The performance of MRIReco.jl is compared to the  Berkeley Advanced Reconstruction Toolbox (BART) and we show that the Julia framework achieves comparable reconstruction speed as the popular C/C++ library.\\
\textbf{Conclusion:}\\
Modern programming languages can bridge the gap between high performance and accessible implementations. MRIReco.jl leverages this fact and contributes a promising environment for future algorithmic development in MRI reconstruction.

\end{abstract}

\section{Introduction}
\label{sec:Intro}
Magnetic resonance imaging (MRI) is a radiation-free tomographic imaging modality allowing for both high spatial resolution and high soft-tissue contrast. This, in combination with the large number of available contrasts, makes it an indispensable tool for many clinical imaging applications. In recent times, acquisition times and spatial resolution have been pushed further through the introduction of new, advanced signal processing techniques such as compressed sensing (CS) \cite{lustig2007sparse,lustig2010spirit} or structured matrix completion \cite{jin2016aloha,haldar2013loraks,shin2014sake}. In a similar way, algorithmic developments have stimulated the development of new techniques for the quantitative imaging of tissue parameters, such as relaxation times, magnetic susceptibility or apparent diffusion coefficients \cite{doneva2010cspm, zhang2015llr,schweser2012heidi,mani2017mussels}. An important catalyst for this development is the availability of open source software. The latter facilitates the development of new methods while taking away the need of a full, self-implemented MRI signal processing pipeline.

An important aspect to keep in mind is that improvements in image quality and / or reduction in scan time are often only possible by computationally intensive algorithms shifting the bottleneck in latency from acquisition to reconstruction. To reduce the reconstruction time to a manageable level massive parallelization including the usage of GPGPU (General Purpose Graphical Purpose Units) \cite{stone2008accelerating,sorensen2009real} hardware acceleration are often necessary. While this allows for acceleration of MRI reconstruction by a factor of 10-1000, it at the time increases the complexity of the software system.

Researchers developing new image reconstruction algorithms usually face two conflicting goals. On the one hand, one seeks a programming environment allowing for an easy translation of mathematical notation into programming code. On the other hand, the program should run as fast as possible. The first goal can be accomplished by using a high-level (HL) programming environment, albeit at the cost of performance. On the other hand, computation speed can be ensured by working in a low-level (LL) programming environment, which results in an increased system complexity. For this reason, one can observe that HL languages, such as Matlab and Python/Numpy \cite{van2011numpy}, are popularly used by mathematical/theoretical researchers. 
In contrast, the LL approach is often used by the vendor of an imaging system, since the product needs to runs in a sufficiently short reconstruction time. In that case, the increase in system complexity is addressed by increasing the development resources.

A popular approach to reconcile both goals is to implement only hot loops in a LL language, while using a HL language for the remaining part. Popular examples for this hybrid approach are Matlab with its Mex interface, i.e. a combination of Matlab and C/C++ and Python with its different ways to call C code (e.g. ctypes). Note that the hybrid approach is not restricted to user code but that entire frameworks are built in this way. Examples include Numpy \cite{van2011numpy}, which is mainly written in C, and machine learning frameworks such as TensorFlow \cite{abadi2016tensorflow} and PyTorch \cite{paszke2019pytorch}, which are implemented in C/C++ with Python language bindings. 

The hybrid approach has many advantages but it also has issues, which  the authors in \cite{bezanson2017julia} summarize in the \textit{two language problem}:
\begin{itemize}
\item The LL implementation is capsuled away from the HL user interface. This is fine in most situations but becomes problematic if custom algorithms need to be implemented.
\item Bridging from one language to the other often has a high computational cost and only amortizes for large problem sizes. Thus, code is often vectorized to minimize the number of calls into the LL language.
\end{itemize}
Beyond that it should be noted that two-language solutions are more complex to set up since multiple compilers and a proper management of dependencies are needed. In consequence, using an existing framework is easy whereas developing new methods/algorithms can be challenging as this often requires programming skills in the LL language used. 

In the realm of MRI, the hybrid approach is very popular. Prominent examples are the packages Gadgetron \cite{hansen2013gadgetron} and BART \cite{uecker2015berkeley}, which both have many features and follow the two-language principle with a core written in C/C++ and bindings to different HL languages. In contrast, SigPy \cite{ong2019sigpy} is a HL package entirely written in Python and allows for rapid prototyping of MRI reconstruction algorithms.

In this paper, we present an MRI reconstruction framework, which yields high performance while allowing for easy prototyping of new algorithms. The framework is implemented in a single programming language, thus avoiding the two-language problem. To achieve this, our framework uses the programming language Julia, which was invented by researchers at the MIT in 2012 with the aim of solving aforementioned two-language problem \cite{bezanson2017julia}. The aim of this paper is two-fold. First, we describe our framework with its main components and we outline how it can be used for the development of new algorithms. Second, we perform comparisons in order to illustrate that the computation speed of our framework indeed matches that of popular frameworks such as BART.

\section{Methods}

When developing an MRI reconstruction framework, it is beneficial to formulate design goals to guide the design. For instance, the developers of Gadgetron formulated the following list of properties, which an MRI reconstruction framework should fulfill:
\begin{enumerate}
\item Free / open
\item Modular
\item Flexible
\item Cross platform
\item High performance
\item Facilitate prototyping
\item Facilitate deployment
\end{enumerate}
For a proper definition of each of the design goals we refer the reader to \cite{hansen2013gadgetron}. \textit{MRIReco.jl} has the same design goals although the deployment aspect is currently not yet addressed. Instead \textit{MRIReco.jl} has the following additional design goals:
\begin{enumerate}
    \item[8.] \textit{Reuse of components}: The goal is to use as many existing software components as possible if and only if they are suitable for the task and appropriately maintained. This includes interdisciplinary standard tools such as the FFT and the conjugated gradients method (CG). This design goal not only implies reusing existing software components but also to put non-MRI specific functionality into dedicated packages if such a package does not yet exist.
    \item[9.] \textit{Hackability}: It should be possible for software developers to access low-level parts of the software and develop extensions or even make modifications to the source code easily. \textit{Hackability} only slightly overlaps with \textit{prototyping} mentioned in \cite{hansen2013gadgetron}. In particular, it means that the gap between being a user and being a developer of a software framework is small.
    \item[10.] \textit{Accessibility}: The software components should be easy to install with only few instructions. It should be simple to access and modify the existing code. 
    \item[11.] \textit{Testing}: The code should be properly tested using continuous integration services.
\end{enumerate}
The last two design goals are certainly also followed by Gadgetron and BART. We note, however, that the two-language principle imposes limits on how well the first two of the additional design goals can be met. For instance, hackability is difficult to address in a two language solution, since code modifications often require knowledge of both the HL and the LL language used.

To make design goal 8 more concrete we note that image reconstruction frameworks often share a large number of generic building blocks. These components include common transformations such as the FFT, standard linear algebra tools, and optimization algorithms to solve the reconstruction problem at hand. In order to share these building blocks it is advantageous to put them into dedicated software packages. This allows developers to reuse building blocks implemented by experts of the given field at hand and it reduces potential errors that could arise when self-implementing them. Finally, we note that the reuse of standard libraries also leads to a higher level of scrutiny for the latter. In turn, this decreases the likelihood of persistent bugs.

\subsection{Julia}
\label{subsec:julia}
To meet the design goals outlined in the previous section, we focus on a new programming language / environment called Julia. Julia was invented by researchers at the MIT in 2012 \cite{bezanson2017julia} and has the aim of solving the two-language problem. Its most important features include:
\begin{itemize}
    \item Julia uses a just-in-time (JIT) compiler based on the LLVM compiler infrastructure \cite{lattner2004llvm} and is therefore capable of generating efficient machine code.
    \item Julia has a sophisticated type system including type inference \cite{milner1978theory}, which allows the compiler to identify statically typed code fragments and in turn emit efficient machine code.
    \item Julia code can be both HL and LL. While hot code paths should be kept type stable for maximizing performance, dynamic programming can be used where convenient code is more important than high performance.
    \item The syntax of Julia is very similar to Matlab and thus familiar to a wide range of researchers.
    \item Julia can call C/C++ code with no overhead.
\end{itemize}
For more details on Julia we refer the reader to Ref. \cite{bezanson2017julia}. 

Julia was designed from the ground up to be capable of generating efficient machine code. In combination with the HL features of the language, this allows us to develop a hackable framework (design goal 9) while retaining a high performance. In contrast, many dynamically typed languages like Python are challenging to JIT compile since they  contain language features preventing  emitting  efficient machine code. Therefore, one of the most successful Python JIT compilers Numba \cite{lam2015numba} requires the user to manually annotate the code to be compiled, and only a subset of the language can be used. Here, again, performance is achieved by enabling the system to determine the types of all variables in a function body. A more general JIT compiler for Python not requiring code annotations is PyPy \cite{bolz2009tracing}.

Another feature of Julia is its built-in package manager. The latter can take care of all the dependencies of a code package. As a consequence, Julia packages are considered to be cheap and it is possible to achieve a fine grained modularization across packages. This modularization is pushed by a large community of package developers, which enrich the power of the Julia programming environment substantially. By exploiting this aspect of the Julia programming environment, \textit{MRIReco.jl} manages to fulfill design goal 8 (reuse of components). Beyond that the clear versioning performed by the julia package manager can greatly simplify reproducible research. This stands in contrast to other approaches, where it is in the hand of the user to download the correct versions of the dependencies at hand.

\subsection{Functionality}

\textit{MRIReco.jl} offers a wide range of functionality while concentrating on basic building blocks that have proven useful for MRI image reconstruction. In particular \textit{MRIReco.jl} currently offers:
\begin{itemize}
    \item \textit{Simulation:} Methods for simulating MRI data based on software phantoms. This includes support for modelling $B_0$ inhomogeneity, $R_2^*$ relaxation and multiple receive coils. Moreover, basic support for simulating multi-echo sequences is implemented. Simulation can be performed using a direct evaluation of the imaging equation or using common approximations such as the NFFT.
    \item \textit{Imaging Operators:} Basic Cartesian and non-Cartesian imaging operators based on FFT and NFFT. Off-resonance and relaxation term aware imaging operators are available as well. These can be evaluated using both data-driven \cite{fessler2007model} and analytical \cite{eggers2007field,knopp2009iterative} approximations. All operators can be combined to form encoding operators for more extended acquisitions such as parallel imaging \cite{pruessmann1999sense} and multi-echo data acquisition.
    \item \textit{Coil Estimation:} \textit{MRIReco.jl} implements methods for determining sensitivity maps including the sum-of-squares method and ESPIRiT \cite{uecker2014espirit}.
    \item \textit{Iterative Reconstruction:} Iterative reconstruction algorithms using solvers such as CGNR \cite{hestenes1952methods}, FISTA \cite{beck2009fista}, ADMM \cite{parikh2014proximal}, and the split Bregman method \cite{goldstein2009split} are available. For regularization, \textit{MRIReco.jl} offers TV, $\ell^1$, $\ell^2$ priors including wavelet sparsification. Density compensation weighting is available as a method for speeding up convergence of iterative solvers \cite{pruessmann2001advances, knopp2007note}. Sampling density weights can be computed for arbitrary trajectories based on the method described in \cite{pipe1999sampling}.
\end{itemize}

\noindent \textit{MRIReco.jl} is implemented in a Julia idiomatic way and uses multiple dispatch to execute different code paths depending on the type of function arguments. We note that this is a similar form of polymorphism known in classical object-orientated languages but that the dispatch is more generic. 
Since julia allows for dynamic programming, it was not necessary to introduce a further abstraction mechanism like a pipeline architecture \cite{hansen2013gadgetron}.

\begin{figure}
    \centering
    \includegraphics{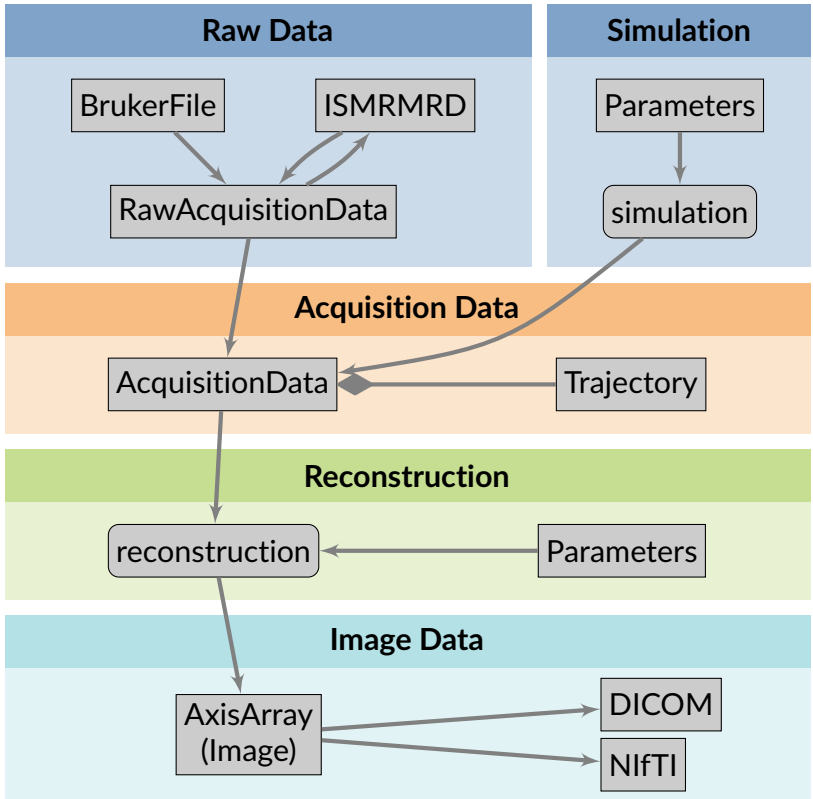}
    
    \caption{Overview of the internal architecture and the typical data flow in \textit{MRIReco.jl}. An \texttt{AcquisitionData} object can be constructed either from a file through a \texttt{RawAcquisitionData} object or using \texttt{simulation}. The \texttt{AcquisitionData} object and additional reconstruction parameters are passed to the \texttt{reconstruction} method that yields an \texttt{AxisArray} object that can be stored in DICOM or NIfTI files.}
    \label{fig:AcqObj}
\end{figure}

An overview of the internal architecture and a typical data flow within \textit{MRIReco.jl} is given in Fig.\,\ref{fig:AcqObj}. In the following sections we sketch individual parts of the framework using small code snippets to illustrate some of the design principles. For more detailed information we refer to the documentation and the source code of the package.

\subsection{Datatypes}
In order to efficiently work with MRI data \textit{MRIReco.jl} introduces two distinct datatypes for its represention. \texttt{RawAcquisitionData} describes the data as it is stored in a data file. Since this data is typically not stored in a form suitable for reconstruction, it is first converted into the type \texttt{AcquisitionData}, describing the pre-processed data. The latter can then be passed to the reconstruction method in order to obtain an image. \texttt{Trajectory} is another important datatype, which is used for describing the MRI sampling trajectory. We outline each datatype in more detail next.
\begin{figure}
    \centering
    \includegraphics{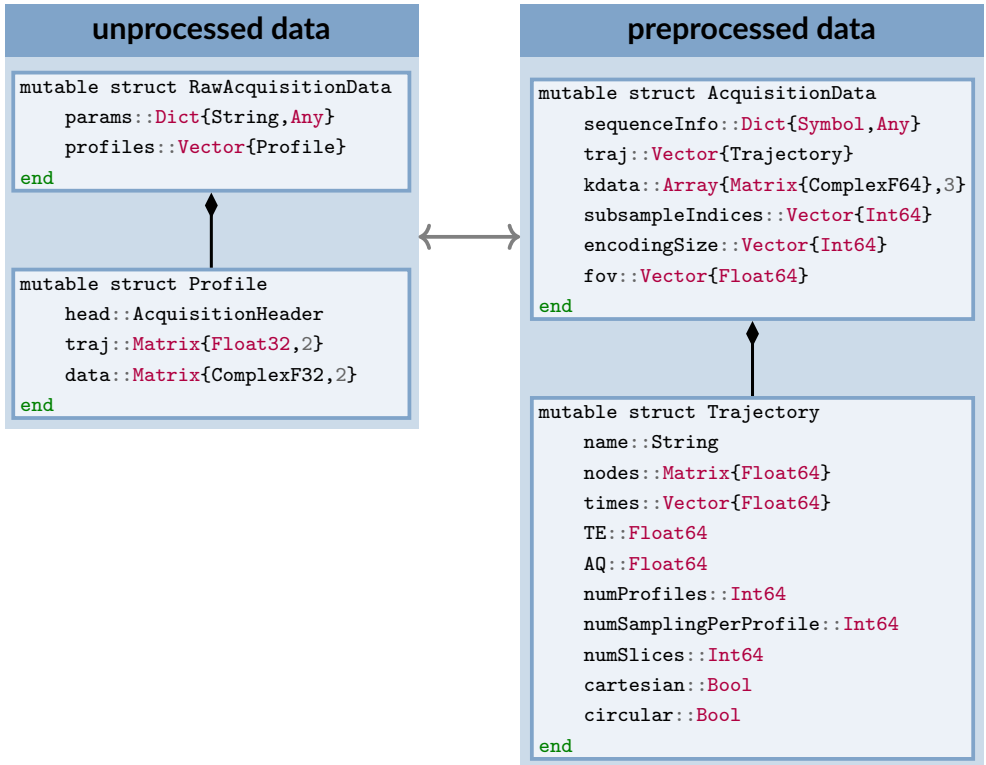}
    \caption{Datatypes used to represent MRI data. The left-hand side contains the julia definition of the datatype \texttt{RawAcquisitionData} used for storing unprocessed MRI data. The right-hand side contains the corresponding definition of the type \texttt{AcquisitionData} used for describing the pre-processed MRI data.}
    \label{fig:acqData}
\end{figure}

\subsubsection{Raw Data}
\texttt{RawAcquisitionData} is a datatype closely resembling the ISMRMRD data format \cite{inati2017ismrm}. Its julia definition is contained in the left-hand side of Fig. \ref{fig:acqData}.
The syntax \texttt{profiles::Vector\{Profile\}} means that the field \texttt{profiles} has the type \texttt{Vector\{Profile\}}. The philosophy of the ISMRMRD format is to store all global metadata in an XML header. This header has a static structure which can be extended by custom fields. \texttt{RawAcquisitionData} uses the \texttt{Dict} \texttt{params} (i.e. an associative array) to store all data that are present in the XML header of an ISMRMRD file. If possible the content of each parameter is converted to an appropriate Julia datatype. For convenience, the structure of the ISMRMRD header is flattened, since this makes all parameters directly accessible.  For instance the parameter 
\begin{minted}[mathescape, %linenos, 
numbersep=5pt, gobble=0, frame=lines, framesep=2mm]{xml}
<reconSpace>
  <matrixSize>
    <x>128</x>
    <y>128</y>
    <z>128</z>
  </matrixSize>
</reconSpace>
\end{minted}
is available in the length-3 vector \texttt{head["reconSize"]}.

Each measurement profile is stored in the type \texttt{Profile}. It  describes the data measured after a single excitation during an MRI experiment. It has members \texttt{head}, \texttt{traj}, and \texttt{data}, which exactly correspond to the structures specified by the ISMRMRD file format. The members of the \texttt{Profile} datatype are also bit-compatible with corresponding HDF5 structs in an ISMRMRD file.

\subsubsection{Pre-Processed Data}
The goal of the \texttt{RawAcquisitionData} datatype is to have a very flexible representation for storing a great variety of measurements. As a matter of fact, this generic representation can be inconvenient when used for reconstruction. For instance, the profiles can be stored in a different order than required for image reconstruction.
For this reason, \textit{MRIReco.jl} uses an additional datatype \texttt{AcquisitionData} to store the data in a way convenient for reconstruction. The  definition of \texttt{AcquisitionData} can be found on the right hand side of Fig. \ref{fig:acqData}.
It contains the sequence information stored in a dictionary, the $k$-space trajectory, the $k$-space data,  several parameters describing the dimensionality of the data, and some additional index vectors.
The $k$-space data \texttt{kdata} has three dimensions encoding
\begin{enumerate}
    \item contrasts/echoes
    \item slices
    \item repetitions
\end{enumerate}
Each element is a matrix whose dimensions encode
\begin{enumerate}
    \item $k$-space nodes
    \item channels/coils
\end{enumerate}
For undersampled data, the indices of the measured samples are stored in the field \texttt{subsampleIndices}. We note that both the \texttt{traj} and the \texttt{subsampleIndices} fields are defined as vectors with one entry for each contrast/echo.

The encoded space is stored in the field \texttt{encodingSize}. It is especially relevant for non-Cartesian trajectories where it is not clear upfront, how large the grid size for reconstruction should be chosen. Finally, the parameter \texttt{fov} describes the physical size of the encoding grid.

\subsubsection{Trajectory}
\begin{figure}
    \centering
    \includegraphics{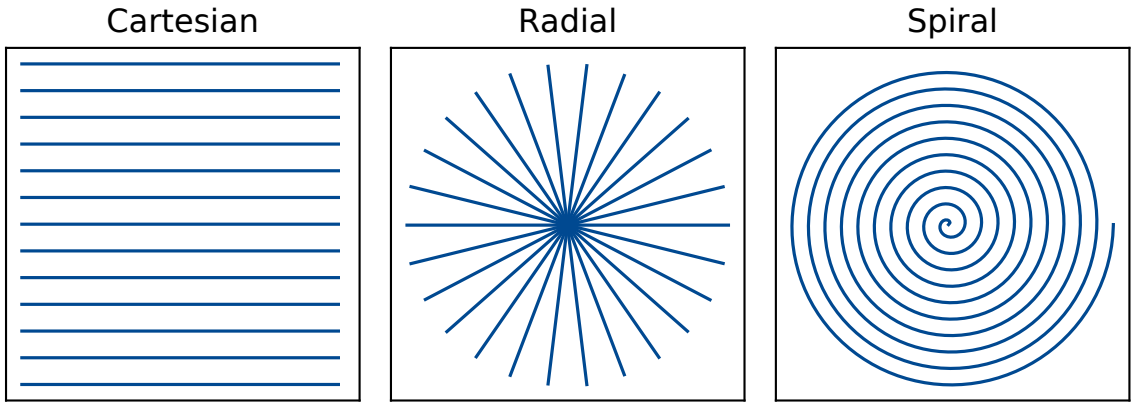}
    \caption{Exemplary trajectories available in  \textit{MRIReco.jl}.}
    \label{fig:trajectory}
\end{figure}

The type \texttt{Trajectory} describes the sampling trajectory of the imaging sequence. Its definition is also contained on the right-hand side of Fig. \ref{fig:acqData}.
Most importantly, the field \texttt{nodes} contains the sampling points as a matrix where the first dimension encodes the $k$-space dimensionality and the second dimension encodes the number of sampling points.
This structure allows implementing arbitrary sampling trajectories. Moreover, \textit{MRIReco.jl} provides constructors for the most common types of trajectories. These include Cartesian, radial, spiral, dual density spiral, variable density spiral, and perturbed spiral trajectories. Beside 2D trajectories, the framework also implements 3D trajectories like the kooshball or the stack of stars trajectory. A selection of exemplary trajectories is illustrated in Fig.~\ref{fig:trajectory}.

\subsubsection{Image Data}

\textit{MRIReco.jl} returns image reconstruction results in the form of an \texttt{AxisArray}, which is special datatype allowing to encode dimensions of an array. For image processing of reconstructed data one can use the package \textit{Images.jl}. Storage of this data is possible using \textit{NIfTi.jl}, \textit{DICOM.jl} or \textit{HDF5.jl}.

\subsection{Raw Data File Handling}
The file handling in \textit{MRIReco.jl} is build around the ISMRMRD file format for which full read and write support is implemented. In addition, a file reader for proprietary files from the vendor Bruker is implemented. The following code example shows how to convert a Bruker MRI file into an ISMRMRD file:
\begin{minted}[mathescape, %linenos, 
numbersep=5pt, gobble=0, frame=lines, framesep=2mm]{julia}
f = BrukerFile(filenameBruker) # create file handle
raw = RawAcquisitionData(f)    # load the data
fout = ISMRMRDFile("outputfile.h5") # create file handle
save(fout, raw) # store the data in the ISMRMRD file 
\end{minted}
We note that the raw data object \texttt{raw} can also be created by simulation. It is thus possible to cache simulated data, which is useful to save computation time associated with more complex simulations.

\subsection{Reconstruction Building Blocks}
In general, MRI image reconstruction aims to recover a discrete image of the transverse magnetization $\zb m\in\mathbb{C}^N$ from a given set of measurements $\zb s\in\mathbb{C}^{M}$ and a signal encoding model described by a linear operator $\zb H\in\mathbb{C}^{M\times N}$. Thus, one seeks a solution to an inverse problem
\begin{equation} \label{Eq:RecoProblem}
\underset{\zbs m}{\operatorname{argmin}} \parallel \zb s - \zb H \zb m \parallel_2^2 + \mathcal{R}(\zb m ), 
\end{equation}
where $\mathcal{R}$ denotes a regularization function expressing prior knowledge about the solution. As a consequence most image reconstruction schemes can be formulated using a set of basic building blocks
\begin{itemize}
    \item \textit{Linear Operators:} These are used to describe the signal encoding operator $\zb H$. Moreover, they describe transformations applied to $\zb m$ within the regularization term $\mathcal{R}(\zb m)$ (e.g. sparsifying transforms used in CS).
    \item \textit{Solvers:} Optimization algorithms used to solve problem \eqref{Eq:RecoProblem}.
    \item \textit{Proximal Maps:} These are associated with the given regularization functions.
\end{itemize}
The interface in \textit{MRIReco.jl} is designed so that all relevant parameters can be passed to the reconstruction method via a dictionary. The reconstruction method uses these parameters to form the main building blocks and solves the corresponding image reconstruction problem. In the following sections, we provide more information on the use and implementation of aforementioned building blocks.

\subsubsection{Linear Operators}
Linear operators are used in multiple places of a reconstruction pipeline. Most importantly, the signal encoding operator $\zb H$ is a discrete approximation of the underlying signal model 
\begin{equation} \label{Eq:ImagingEquation}
s_p(t) = \int_{\mathbb{R}^d} c_p(\zb r) m(\zb r) {\rm e}^{- z(\zbs r)t} {\rm e}^{-2\pi \ti \zbs k(t) \cdot \zbs r } \d \zb r.
\end{equation}
Here $s_p(t)$ denotes the demodulated signal received in the $p$-th of $P$ receive coils at time $t$. Moreover, $m(\zb r)$ is the transverse magnetization of the object at position $\zb r$ and $c_p(\zb r)$ are the receive coil sensitivities. Finally, the term $z(\zb r)\in \mathbb{C}$ contains the $R_2^*$ and $B_0$ maps in its respective real and imaginary parts. 
As a second application, linear operators can be applied to $\zb m$ in the regularizer $\mathcal{R}$. This commonly happens in compressed-sensing-type reconstructions, where the image needs to be transformed into a sparse representation. 

All operators are implemented in a matrix-free manner. This means that they are characterized solely by their action when applied to a vector. This allows to evaluate operators using efficient algorithms, such as the FFT and NFFT, while avoiding storage of the underlying matrix representation. Other common matrix operations are implemented in complete analogy. For instance, one can apply a linear operator using the \texttt{*}-operator or form its adjoint using the postfix \texttt{'}. Similarly, linear operators can be composed using either of the operators \texttt{*} and \texttt{$\circ$}, and their size can be determined using the \texttt{size}-function.\footnote{We note that the term operator is used here for both the binary mathematical operation and the linear mapping.}  

\textit{MRIReco.jl} provides implementations of the operators commonly used for MRI image reconstruction. These include
\begin{itemize}
\item \texttt{FFTOp}: A multidimensional FFT operator.
\item \texttt{NFFTOp}: A multidimensional NFFT operator.
\item \texttt{FieldmapNFFTOp}: An extension of the NFFT operator taking into account complex fieldmaps.
\item \texttt{SensitivityOp}: An operator, which multiples an image by a set of coil-sensitivities as used in SENSE-type reconstructions.
\item \texttt{SamplingOp}: An operator for (sub)sampling ($k$-space) data.
\item \texttt{WeightingOp}: A weighting operator used for tasks such as sampling density compensation.
\end{itemize}
Additional operators, such as the Wavelet transform and finite differences operators, are reexported from the Julia package \textit{SparsityOperators.jl}.

In addition to theses methods, \textit{MRIReco.jl} provides high-level constructors that compose aforementioned operators and return signal encoding operators for the different reconstruction schemes. These are automatically called by the reconstruction methods implemented.
Alternatively, each operator can be built manually by calling the corresponding constructor. In this way the preimplemented operators can be used as building blocks when developing new algorithms.

Finally, we note that iterative solvers often require repeated application of the normal operator $\zb N = \zb H^{H}\zb H$ of the encoding operator. Thus, algorithms can sometimes be accelerated by optimizing the normal operator instead of optimizing the encoding operator itself. For instance, one can exploit the Toeplitz structure of $\zb H^{H}\zb H$ when $\zb H$ is an NFFT. To allow for this kind of optimization, \textit{MRIReco.jl} reexports the type \texttt{normalOperator} from the packages \textit{SparsityOperators.jl} and \textit{RegularizedLeastSquares.jl}. When an optimized implementation of the normal operator exists, the latter can be used by simply overloading the constructor function \texttt{normalOperator}.

\subsubsection{Solvers}
In order to solve the reconstruction problem at hand, $\textit{MRIReco.jl}$ uses the infrastructure provided by the package \textit{RegularizedLeastSquares.jl}. The latter implements popular iterative optimization methods, such as the CGNR, FISTA  and ADMM. For all the reconstruction methods implemented in \textit{MRIReco.jl} the solver can be determined by assigning its name to the parameter \texttt{:solver} in the dictionary containing the reconstruction parameters.

\subsubsection{Regularization}
To describe regularization functions, \textit{MRIReco.jl} uses the type \texttt{Regularization} from the package \textit{RegularizedLeastSquares.jl}.
Most notably, this type contains a function to compute the associated proximal map. This approach is very generic in the sense that most common solvers incorporate regularization in the form of a proximal map.

For convenience, \textit{RegularizedLeastSquares.jl} implements several common regularization functions such as TV, $\ell^1$, $\ell^2$ and low rank regularization. Analogously to the solver to be used, these regularization functions can be specified by assigning their respective names to the parameter \texttt{:regularization} (e.g. \texttt{params[:regularization] = "L1"} ). 
Alternatively, one can assign one (or more) \texttt{Regularization} objects to aforementioned parameter. This allows the incorporation of custom regularization functions. In this case, the main work to be done is the implementing the corresponding proximal map. Finally, the preimplemented regularization objects can serve as building blocks when developing new optimization algorithms.

\subsection{High-Level Reconstruction}

\begin{figure}
    \centering
    \includegraphics{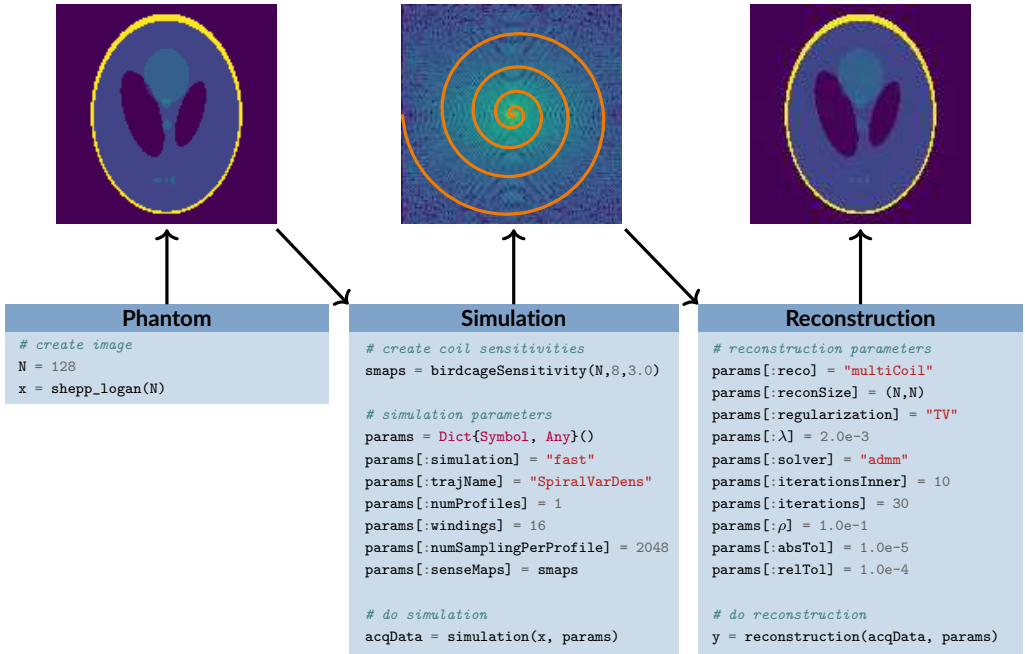}
    \caption{Example of a high-level simulation and reconstruction script using the software package \textit{MRIReco.jl}. In the lower part of the figure the generating code snippets are shown while the resulting data is shown in the top part. The example starts by generating a $128\times 128$ Shepp–Logan  phantom (left column), which is used for simulating 8-fold undersampled $k$-space data (middle column). Finally, a simple $TV$-regularized iterative parallel imaging reconstruction is performed (right column).}
    \label{fig:reconHighLevel}
\end{figure}

Next we sketch a high-level simulation and reconstruction with \textit{MRIReco.jl}. As outlined before, the interface is designed in such a way that all parameters are passed to the routine \texttt{simulation} and \texttt{reconstruction} via a parameter dictionary. This approach is very generic and allows specifying a large set of parameters without using long argument lists.  The dictionary can also be stored in an  XML or TOML file.

\noindent The example simulation uses a $128 \times 128$ pixel sized Shepp-Logan phantom, 8 birdcage coil sensitivities, and a variable density spiral trajectory with 1 interleave and 2048 samples. This corresponds to an 8-fold undersampling. Afterwards, reconstruction is performed using a SENSE-type compressed sensing reconstruction with TV-regularization. The reconstruction problem is then solved using  ADMM with 10 inner CG-iterations and 30 outer iterations. The resulting image is shown on the right-hand side of Fig. \ref{fig:reconHighLevel}.

\subsection{Low-Level Reconstruction}
The high-level reconstruction outlined in the previous section allows performing image reconstruction for a wide range of MRI imaging scenarios. Alternatively, the building blocks in \textit{MRIReco.jl} can be used in a more low-level way to implement custom reconstruction methods. In the next example we illustrate this using the example of a simple gridding reconstruction. With an \texttt{AcquisitionData} object at hand, a gridding reconstruction could be implemented using the following code snippet. First, density compensation weights (\texttt{samplingDensity}) are computed and the trajectory to be used is extracted. After forming the gridding operator (\texttt{NFFTOp}), the reconstructed image is obtained by applying its adjoint to the weighted $k$-space data. This illustrates that the notation within \textit{MRIReco.jl} is very close to the underlying mathematical description.

\begin{minted}[mathescape, %linenos, 
numbersep=5pt, gobble=0, frame=lines, framesep=2mm]{julia}
using MRIReco
  
# load or simulate acqData

# custom low-level reconstruction
reconSize = acqData.encodingSize[1:2]
weights = samplingDensity(acqData, reconSize)[1]
tr = trajectory(acqData)
F = NFFTOp(reconSize, tr)
kdata = kData(acqData,1,1,1) .* (weights.^2)
reco = adjoint(F) * kdata
\end{minted}

\subsection{Parallelization}

One important way to speed up reconstruction is to make use of parallel code execution.
\textit{MRIReco.jl} supports  multi-threading (i.e. shared memory parallelism) as the primary form of parallelism and has also preliminary support for GPU acceleration, which has been recently added to the package \textit{NFFT.jl} (since  version 0.6).
To achieve decent speed-ups irrespective of the reconstruction setting, multi-threading is implemented in different parts of \textit{MRIReco.jl}. On a low level, several imaging operators, such as the multi-coil imaging operator and the fieldmap-aware imaging operator, require the computation of multiple NFFTs, which can be done in parallel. On a higher level, one often wants to perform multiple independent reconstructions. This includes the reconstruction of multiple slices of a 2d acquisition or the reconstruction of multiple independent coil images. Therefore, all loops over independent reconstructions are parallelized as well. Parallelization on different levels is possible without risking oversubscription of CPU cores with too many threads since Julia uses a thread pool that allows for nested parallelism similar to OpenMP \cite{dagum1998openmp}, Cilk \cite{blumofe1996cilk}, and Intel TBB \cite{kukanov2007foundations}.

\subsection{Availability and Platform Support}

\textit{MRIReco.jl} is developed within a public Git repository hosted at Github.\footnote{\url{https://github.com/MagneticResonanceImaging/MRIReco.jl}} The project contains online documentation that can be accessed from the project homepage. Bug reports, feature requests and comments can be made using an issue tracker. Any commit made to \textit{MRIReco.jl} is tested using continuous integration services. \textit{MRIReco.jl} is supposed to run on any operating system and platform that Julia itself supports. Currently, the test suite runs successfully on Linux, OS X, and Windows.

The software package is licensed under the MIT license\footnote{\url{https://opensource.org/licenses/MIT}}, as are most parts of Julia  and its package ecosystem. The MIT license is a  permissive license and allows to use the code even in a closed-source application. \textit{MRIReco.jl} has only one GPL dependency (FFTW \cite{frigo1998fftw}), which would need to be replaced prior inclusion into a closed source application\footnote{Massachusetts Institute of Technology (MIT) and Intel (MKL, Math Kernel Library) provide binary compatible FFTW implementations that can be used in closed-source applications.}. 

\subsection{Experimental Evaluation}

After giving an overview of the functionality and implementation of \textit{MRIReco.jl} we next apply \textit{MRIReco.jl} to openly available MRI data and perform a comparison with an existing MRI reconstruction framework.  

The first test aims at testing/demonstrating the full reconstruction pipeline from loading MRI data in the commonly used ISMRMRD format to performing image reconstruction. For this purpose, we downloaded a publicly available MRI dataset from the database \url{http://mridata.org} \cite{ong2018mridata}. The dataset contains data of a human knee acquired using a 3d FSE sequence and an 8-channel receive coil on a GE scanner.
The data was acquired with a FOV of 160 mm x 160 mm x 124.8 mm, a matrix size of 320 x 274 x 208 and a TR/TE of 1400 ms / 20 ms. The data was measured using variable density Poisson disk sampling with a fully sampled calibration area of size 35x35 and an overall undersampling factor of 7.13. After loading it, the data was converted to 2d data by applying a Fourier transform along the readout direction. Next, coil sensitivy maps were obtained using the ESPIRiT implementation contained in \textit{MRIReco.jl}. Finally, image slices were reconstructed using a SENSE-type reconstruction with $\ell^1$-regularization in the Wavelet domain and a regularization parameter of 0.2. The reconstruction problem was solved using the ADMM with 30 iterations and 10 iterations of the inner CG method.

Secondly, we perfomed a comparison of \textit{MRIReco.jl} with the popular C/C++ reconstruction framework BART \cite{uecker2015berkeley}. As a model problem we chose an iterative SENSE reconstruction using the radial dataset published as part of the ISMRM reproducibility challenge 1 \cite{maier2020cgsense}. 
It consists of a brain dataset acquired with 12 coils using a radial trajectory with 96 profiles and 512 samples per profile. The coil sensitivities are estimated from the data itself using the ESPIRiT implementation that is part of each framework. As proposed in the reproducibility challenge, the fully sampled dataset was retrospectively undersampled by reduction factors between $R=1$ and $R=4$. For both frameworks, images were recovered using an iterative $\ell^2$-regularized SENSE reconstruction based on the CG method. In all cases we used 20 iterations and a regularization parameter of $0.01$. 
It is run on a workstation equipped with two AMD EPYC 7702 CPUs running at 2.0 GHz (256 cores in total) and a main memory of 1024 GB.

The data/software that support the findings of this study are openly available in \textit{MRIReco.jl} at \url{10.5281/zenodo.4464857}, SHA-1 hash 72fbbd0, and \textit{MRIRecoBenchmarks} at \url{10.5281/zenodo.4467979}, SHA-1 hash b24b60c. To be precise, the reconstruction of the knee dataset is contained in the examples folder in \textit{MRIReco.jl}, while the code for the comparison with BART is contained in \textit{MRIRecoBenchmarks}.

\section{Results}

\subsection{Reconstruction Results of the Knee Data Set}
\begin{figure}
     \centering
    \includegraphics{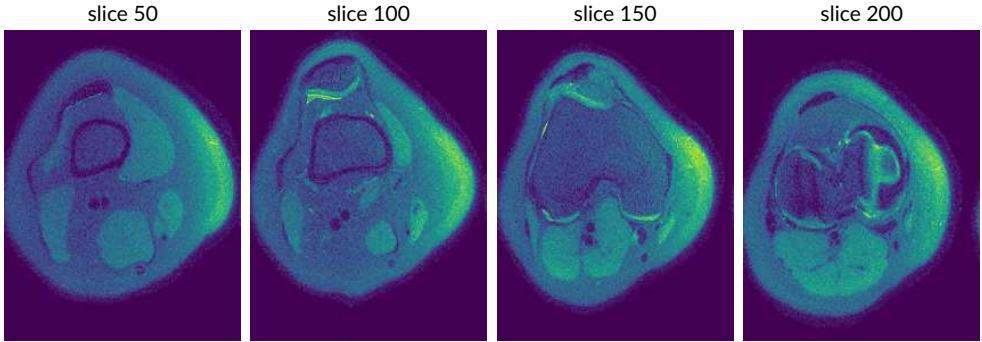}
    \caption{Results of the $\ell^1$-Wavelet reconstruction of the knee data set. The reconstructed images are shown for the slices 50, 100, 150 and 200 (along readout direction).}
    \label{fig:kneeReco}
\end{figure}
Fig.~\ref{fig:kneeReco} shows images of the reconstructed knee dataset for some exemplary slices. These results illustrate a typical application for researchers, who wish to test their reconstruction methods not only using simulation data but also data from MRI scanners. As illustrated in the example code accessible in the Git repository, this can by achieved quite easily with \textit{MRIReco.jl} in conjunction with the ISMRMRD format. Thus, all that is needed to work with raw data is to convert the latter from its vendor specific format into the ISMRMRD format. Afterwards, the data can be reconstructed using either one of the preimplemented reconstruction methods or by a custom reconstruction method making use of the low-level building blocks provided by \textit{MRIReco.jl}.

\subsection{Runtime Performance}

Designing a proper performance benchmark for independent software frameworks can be a challenging task since frameworks  often differ not only in the implementation but also in the choice of implemented algorithms/optimizations and in the choice of default reconstruction parameters. 
Since both BART and \textit{MRIReco.jl} implement multi-threading we performed benchmarks for different numbers of threads (1,4,8,12). For each framework, the reconstruction is performed multiple times and the minimum time is used for comparison. In this way, both frameworks are benchmarked under idealized but comparable conditions. In particular this procedure considers hot CPU caches.

BART by default uses the Toeplitz optimization for efficient multiplication with the normal matrix. During the development of the benchmark we implemented this feature as well to match the implementation in BART. In addition to the Toeplitz optimization, which implies using FFTs with an oversampling of factor $\sigma=2.0$, \textit{MRIReco.jl} also allows running the code without the Toeplitz optimization but with a smaller oversampling factor such as $\sigma=1.25$ (see \cite{eggers2002comparison} for investigation of oversampling factor sizes). Although this may slightly reduce the accuracy of the NFFT approximation, we observe that in practice the approximation error is so small that it is not visually perceptible in the reconstructed images.

\begin{figure}
    \centering
    \includegraphics{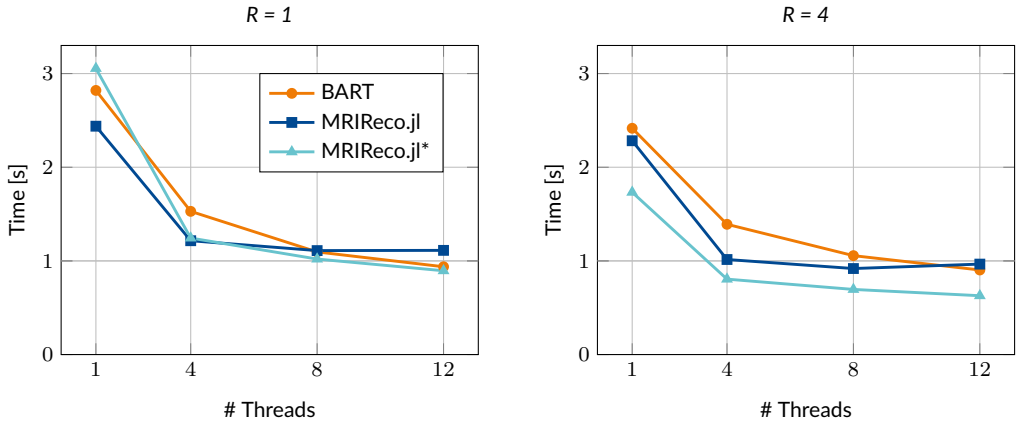}
    \caption{Performance comparison between BART and \textit{MRIReco.jl} shown are the minimum reconstruction times of an iterative SENSE reconstruction for different numbers of threads ranging from $1$ to $12$. On the right, the results for reduction factor $R=4$ are shown while on the left, the results without data reduction can be seen. Orange shows the reconstruction times for BART while dark blue shows the reconstruction times for \textit{MRIReco.jl} both using the Toeplitz optimization. Light blue shows reconstruction time for \textit{MRIReco.jl} without the Toeplitz approach using an oversampling factor $\sigma=1.25$.}
    \label{fig:performance}
\end{figure}

The results of the performance comparison are summarized in Fig.~\ref{fig:performance} for the reduction factors $R=1$ and $R=4$. One can see that both reconstruction frameworks achieve very similar reconstruction times. \textit{MRIReco.jl} is slightly faster for 1 and 4 threads while BART is faster for 12 threads. When comparing the result with and without Toeplitz optimization one can see that for $R=1$ both approaches achieve similar performance while for $R=4$ the non-Toeplitz reconstruction with oversampling factor $\sigma=1.25$ is clearly faster. This can be explained by the smaller size of the FFT and the fact that for $R=4$ significantly fewer points need to be gridded compared to the case where $R=1$.

\subsection{Reconstruction Accuracy}

\begin{figure}
    \centering
    \includegraphics{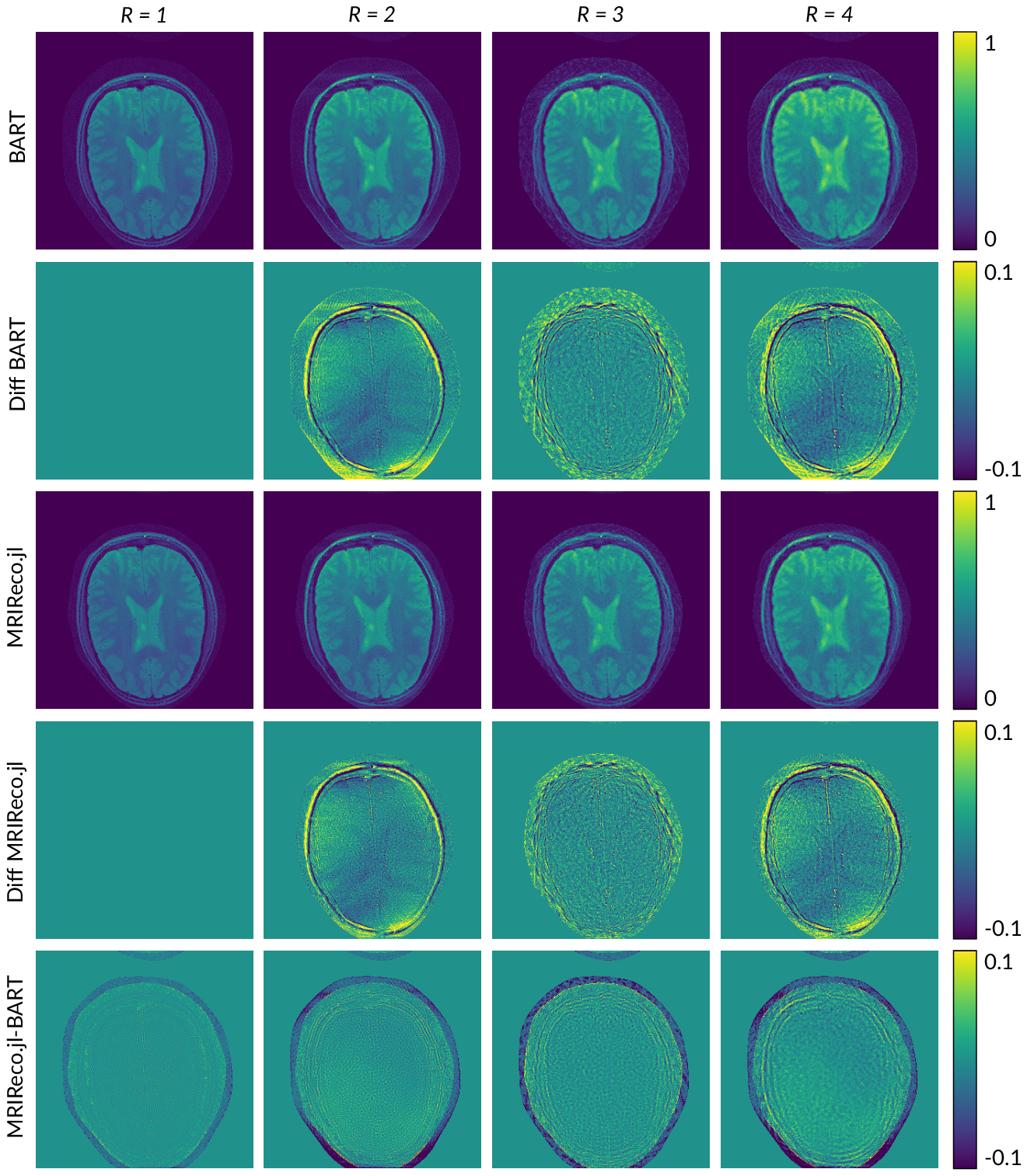}    
    
    \caption{Comparison of SENSE reconstructions of a public brain imaging dataset using BART and \textit{MRIReco.jl}. The reconstruction results are shown in rows one and three, while rows two and four show difference maps to the reference reconstruction ($R=1$). Row five shows difference maps between the images reconstructed using both frameworks for the same reduction factor.}
    \label{fig:comparisonImages}
\end{figure}

Reconstruction results are summarized in Fig.~\ref{fig:comparisonImages}. They are shown for reduction factors $R=1,2,3,4$ using the same reconstruction parameters. For \textit{MRIReco.jl} the images  obtained using the Toeplitz optimization are shown. For both frameworks only a minor decrease of image quality can be observed with increasing reduction factor. When looking at difference maps between reduction factor $R=1$ and higher reduction factors (row two and four) one observes that $R=3$ shows a smaller deviation in  outer image regions of the head than reduction factors $R=2$ and $R=4$.

When comparing the reconstruction results of the two frameworks one can hardly see a difference. Only the outer regions beyond the head look slightly different, which is likely caused by differences in the coil estimation algorithm. Even the difference maps compared to the $R=1$ reconstruction look very similar, which indicates that both frameworks implement the iterative SENSE reconstruction in a similar way. This is further supported by the difference maps between the two reconstruction frameworks shown in the fifth row. These show mostly noise in the head region.

\subsection{Syntactic Comparison}
\begin{table}
\centering
\begin{tabular}{ |c|c|c| } 
 \hline
 task & BART & MRIReco.jl \\
 \hline
 data loading \& conversion & 5 & 7 \\
 gridding \& ESPIRiT & 4 & 2 \\ 
 undersampling data & 3 & 1 \\
 reconstruction parameters & 0 & 10 \\
 reconstruction & 1 & 1 \\
 sum & 13 & 21 \\
 \hline
\end{tabular}
\caption{Number of code lines needed to perform the central tasks associated with the reconstruction of the brain dataset.}
\label{tab1:UserLines}
\end{table}

In order to provide a rough estimate for the complexity of the user interface, Table \ref{tab1:UserLines} summarizes the number of code lines required to perform the central tasks associated with the reconstruction of the brain dataset. This count excludes code lines specific to the benchmarking, such as setting up the benchmark or running multiple trials of the reconstruction. The results show that both frameworks require a similar number of code lines to perform  reconstruction. The main difference arises from the fact that the MRIReco.jl-implementation uses a dictionary to store reconstruction parameters, which is not required for BART. We note however that passing the parameters as a dictionary is done solely for the purpose of keeping the code readable. Alternatively, the parameters could be passed directly as keyword arguments as done by BART. We conclude that both frameworks perform image reconstruction with a comparably complex interface and we acknowledge that the complexity of the final code depends on the preferences of the user at hand.

\section{Discussion}

\textit{MRIReco.jl} started as an experiment of the authors to check the suitability of Julia for developing an MRI reconstruction framework. Initial success was measured in a rapid development experience while still generating programs that can compete in terms of runtime speed with equivalent C/C++ programs. From this point the framework was developed to make it not only usable for the authors themselves but also for other users. We characterize the status of the project as usable but not yet finished. This implies that interfaces of upcoming versions might slightly change and that parts of the documentation are still incomplete.

In a performance benchmark it was shown that \textit{MRIReco.jl} achieves similar and sometimes even better performance than the state-of-the-art C/C++ MRI reconstruction framework BART. We note that the benchmark was performed for a very specific reconstruction algorithm (iterative SENSE) and its results are not directly applicable to other aspects of each of the frameworks.

One key philosophy of \textit{MRIReco.jl} is to reuse existing code and keep the code base small. This makes the package more maintainable as it avoids code duplication and keeps the responsibility for code in the original software repositories. Despite its advantages, this approach requires a package management system capable of handling complex version dependencies. Julia readily addresses this aspect with its integrated package manager. An important responsibility in a package with many dependencies is to maintain compatibility with the depending packages. This can increase the complexity if the packages are not managed by oneself. Fortunately, the Julia dependency system allows to pin packages to certain versions such that breaking API changes in depending packages can be avoided.

One weak point of \textit{MRIReco.jl} is that no functionality for deployment on an MRI scanner is implemented yet. 
There are basically three different implementations thinkable. The first is to write a server in Julia and communicate with the host server via TCP/IP, as was done in \cite{graser2019human}. The second possibility is to embed \textit{MRIReco.jl} into an existing C/C++ program. In this way it would be possible to integrate \textit{MRIReco.jl} into for instance Gadgetron and implement a \textit{JuliaGadget} similar to the existing \texttt{PythonGadget}. Since Julia arrays have a C-compatible binary format, it is possible to pass data from C to Julia by only passing the pointer to the data. Finally, deployment can be achieved using automated offline reconstruction techniques such as Yarra \cite{Yarra} and Autorec \cite{Autorec}.

Julia itself as a language has evolved since its introduction in 2012 into a stable, featureful language that can compete with more widely developed languages such as Python and C/C++. One of the remaining issues of Julia compared to other programming languages is its latency. Code compilation in Julia is currently done right before code execution and thus can add a certain amount of latency. The latency of packages is reduced to some extent by so-called pre-compilation but the current Julia version (1.5.3) stores only an intermediate representation and not the native machine code. The remaining latency issue can be mitigated by the package \textit{Revise.jl}, which allows to cache compiled code during a julia session. An alternative for deployment usage is the package \textit{PackageCompiler.jl}, which allows to either compile packages into the system image of Julia or to generate standalone executables.

\section{Conclusions}
In conclusion we have introduced a new image reconstruction framework for MRI, which is both performant and accessible. Two key aspects of \textit{MRIReco.jl} are its modularity and its open interface. These make it a useful tool not only for performing image reconstruction but also for the development and testing of new reconstruction methods. The framework is implemented purely in the programming language Julia and reaches similar performance and similar image quality as other popular image reconstruction frameworks that are implemented in a low-level programming language.

\section*{Acknowledgements}

We acknowledge many fruitful discussions with Martin Uecker on the usage of BART and the design of the benchmarks.


\begin{thebibliography}{47}
\providecommand{\natexlab}[1]{#1}
\providecommand{\url}[1]{\texttt{#1}}
\providecommand{\urlprefix}{}

\bibitem[{Lustig et~al.(2007)Lustig, Michael and Donoho, David and Pauly, John
  M}]{lustig2007sparse}
Lustig M, Donoho D, Pauly JM.
\newblock Sparse MRI: The application of compressed sensing for rapid MR
  imaging.
\newblock Magnetic Resonance in Medicine: An Official Journal of the
  International Society for Magnetic Resonance in Medicine
  2007;58(6):1182--1195.

\bibitem[{Lustig and Pauly(2010)Lustig, Michael and Pauly, John
  M}]{lustig2010spirit}
Lustig M, Pauly JM.
\newblock SPIRiT: iterative self-consistent parallel imaging reconstruction
  from arbitrary k-space.
\newblock Magnetic resonance in medicine 2010;64(2):457--471.

\bibitem[{Jin et~al.(2016)Jin, Kyong Hwan and Lee, Dongwook and Ye, Jong
  Chul}]{jin2016aloha}
Jin KH, Lee D, Ye JC.
\newblock A general framework for compressed sensing and parallel MRI using
  annihilating filter based low-rank Hankel matrix.
\newblock IEEE Transactions on Computational Imaging 2016;2(4):480--495.

\bibitem[{Haldar(2013)Haldar, Justin P}]{haldar2013loraks}
Haldar JP.
\newblock Low-rank modeling of local $ k $-space neighborhoods (LORAKS) for
  constrained MRI.
\newblock IEEE transactions on medical imaging 2013;33(3):668--681.

\bibitem[{Shin et~al.(2014)Shin, Peter J and Larson, Peder EZ and Ohliger,
  Michael A and Elad, Michael and Pauly, John M and Vigneron, Daniel B and
  Lustig, Michael}]{shin2014sake}
Shin PJ, Larson PE, Ohliger MA, Elad M, Pauly JM, Vigneron DB, et~al.
\newblock Calibrationless parallel imaging reconstruction based on structured
  low-rank matrix completion.
\newblock Magnetic resonance in medicine 2014;72(4):959--970.

\bibitem[{Doneva et~al.(2010)Doneva, Mariya and B{\"o}rnert, Peter and Eggers,
  Holger and Stehning, Christian and S{\'e}n{\'e}gas, Julien and Mertins,
  Alfred}]{doneva2010cspm}
Doneva M, B{\"o}rnert P, Eggers H, Stehning C, S{\'e}n{\'e}gas J, Mertins A.
\newblock Compressed sensing reconstruction for magnetic resonance parameter
  mapping.
\newblock Magnetic Resonance in Medicine 2010;64(4):1114--1120.

\bibitem[{Zhang et~al.(2015)Zhang, Tao and Pauly, John M and Levesque, Ives
  R}]{zhang2015llr}
Zhang T, Pauly JM, Levesque IR.
\newblock Accelerating parameter mapping with a locally low rank constraint.
\newblock Magnetic resonance in medicine 2015;73(2):655--661.

\bibitem[{Schweser et~al.(2012)Schweser, Ferdinand and Sommer, Karsten and
  Deistung, Andreas and Reichenbach, J{\"u}rgen Rainer}]{schweser2012heidi}
Schweser F, Sommer K, Deistung A, Reichenbach JR.
\newblock Quantitative susceptibility mapping for investigating subtle
  susceptibility variations in the human brain.
\newblock Neuroimage 2012;62(3):2083--2100.

\bibitem[{Mani et~al.(2017)Mani, Merry and Jacob, Mathews and Kelley, Douglas
  and Magnotta, Vincent}]{mani2017mussels}
Mani M, Jacob M, Kelley D, Magnotta V.
\newblock Multi-shot sensitivity-encoded diffusion data recovery using
  structured low-rank matrix completion (MUSSELS).
\newblock Magnetic resonance in medicine 2017;78(2):494--507.

\bibitem[{Stone et~al.(2008)Stone, Sam S and Haldar, Justin P and Tsao,
  Stephanie C and Sutton, BP and Liang, Z-P and others}]{stone2008accelerating}
Stone SS, Haldar JP, Tsao SC, Sutton B, Liang ZP, et~al.
\newblock Accelerating advanced MRI reconstructions on GPUs.
\newblock Journal of parallel and distributed computing 2008;68(10):1307--1318.

\bibitem[{Sorensen et~al.(2009)Sorensen, Thomas Sangild and Atkinson, David and
  Schaeffter, Tobias and Hansen, Michael Schacht}]{sorensen2009real}
Sorensen TS, Atkinson D, Schaeffter T, Hansen MS.
\newblock Real-time reconstruction of sensitivity encoded radial magnetic
  resonance imaging using a graphics processing unit.
\newblock IEEE transactions on medical imaging 2009;28(12):1974--1985.

\bibitem[{Van Der~Walt et~al.(2011)Van Der Walt, Stefan and Colbert, S Chris
  and Varoquaux, Gael}]{van2011numpy}
Van Der~Walt S, Colbert SC, Varoquaux G.
\newblock The NumPy array: a structure for efficient numerical computation.
\newblock Computing in Science \& Engineering 2011;13(2):22.

\bibitem[{Abadi et~al.(2016)Abadi, Mart{\'\i}n and Barham, Paul and Chen,
  Jianmin and Chen, Zhifeng and Davis, Andy and Dean, Jeffrey and Devin,
  Matthieu and Ghemawat, Sanjay and Irving, Geoffrey and Isard, Michael and
  others}]{abadi2016tensorflow}
Abadi M, Barham P, Chen J, Chen Z, Davis A, Dean J, et~al.
\newblock Tensorflow: A system for large-scale machine learning.
\newblock In: 12th $\{$USENIX$\}$ symposium on operating systems design and
  implementation ($\{$OSDI$\}$ 16); 2016. p. 265--283.

\bibitem[{Paszke et~al.(2019)Paszke, Adam and Gross, Sam and Massa, Francisco
  and Lerer, Adam and Bradbury, James and Chanan, Gregory and Killeen, Trevor
  and Lin, Zeming and Gimelshein, Natalia and Antiga, Luca and
  others}]{paszke2019pytorch}
Paszke A, Gross S, Massa F, Lerer A, Bradbury J, Chanan G, et~al.
\newblock Pytorch: An imperative style, high-performance deep learning library.
\newblock In: Advances in neural information processing systems; 2019. p.
  8026--8037.

\bibitem[{Bezanson et~al.(2017)Bezanson, Jeff and Edelman, Alan and Karpinski,
  Stefan and Shah, Viral B}]{bezanson2017julia}
Bezanson J, Edelman A, Karpinski S, Shah VB.
\newblock Julia: A fresh approach to numerical computing.
\newblock SIAM review 2017;59(1):65--98.

\bibitem[{Hansen and S{\o}rensen(2013)Hansen, Michael Schacht and S{\o}rensen,
  Thomas Sangild}]{hansen2013gadgetron}
Hansen MS, S{\o}rensen TS.
\newblock Gadgetron: an open source framework for medical image reconstruction.
\newblock Magnetic resonance in medicine 2013;69(6):1768--1776.

\bibitem[{Uecker et~al.(2015)Uecker, Martin and Ong, Frank and Tamir, Jonathan
  I and Bahri, Dara and Virtue, Patrick and Cheng, Joseph Y and Zhang, Tao and
  Lustig, Michael}]{uecker2015berkeley}
Uecker M, Ong F, Tamir JI, Bahri D, Virtue P, Cheng JY, et~al.
\newblock Berkeley advanced reconstruction toolbox.
\newblock In: Proc. Intl. Soc. Mag. Reson. Med, vol.~23; 2015. p. 2486.

\bibitem[{Ong and Lustig(2019)Ong, Frank and Lustig, Michael}]{ong2019sigpy}
Ong F, Lustig M.
\newblock SigPy: A Python Package for High Performance Iterative
  Reconstruction.
\newblock In: Proc. Intl. Soc. Mag. Reson. Med, vol.~27; 2019. p. 4819.

\bibitem[{Lattner and Adve(2004)Lattner, Chris and Adve,
  Vikram}]{lattner2004llvm}
Lattner C, Adve V.
\newblock LLVM: A compilation framework for lifelong program analysis \&
  transformation.
\newblock In: Proceedings of the international symposium on Code generation and
  optimization: feedback-directed and runtime optimization IEEE Computer
  Society; 2004. p.~75.

\bibitem[{Milner(1978)Milner, Robin}]{milner1978theory}
Milner R.
\newblock A theory of type polymorphism in programming.
\newblock Journal of computer and system sciences 1978;17(3):348--375.

\bibitem[{Lam et~al.(2015)Lam, Siu Kwan and Pitrou, Antoine and Seibert,
  Stanley}]{lam2015numba}
Lam SK, Pitrou A, Seibert S.
\newblock Numba: A llvm-based python jit compiler.
\newblock In: Proceedings of the Second Workshop on the LLVM Compiler
  Infrastructure in HPC ACM; 2015. p.~7.

\bibitem[{Bolz et~al.(2009)Bolz, Carl Friedrich and Cuni, Antonio and
  Fijalkowski, Maciej and Rigo, Armin}]{bolz2009tracing}
Bolz CF, Cuni A, Fijalkowski M, Rigo A.
\newblock Tracing the meta-level: PyPy's tracing JIT compiler.
\newblock In: Proceedings of the 4th workshop on the Implementation,
  Compilation, Optimization of Object-Oriented Languages and Programming
  Systems ACM; 2009. p. 18--25.

\bibitem[{Fessler and Nol(2007)Fessler, Jeffrey A and Nol, Douglas
  C}]{fessler2007model}
Fessler JA, Nol DC.
\newblock Model-based MR image reconstruction with compensation for
  through-plane field inhomogeneity.
\newblock In: 2007 4th IEEE International Symposium on Biomedical Imaging: From
  Nano to Macro IEEE; 2007. p. 920--923.

\bibitem[{Eggers et~al.(2007)Eggers, Holger and Knopp, Tobias and Potts,
  Daniel}]{eggers2007field}
Eggers H, Knopp T, Potts D.
\newblock Field inhomogeneity correction based on gridding reconstruction for
  magnetic resonance imaging.
\newblock IEEE Transactions on Medical Imaging 2007;26(3):374--384.

\bibitem[{Knopp et~al.(2009)Knopp, Tobias and Eggers, Holger and Dahnke, Hannes
  and Prestin, J{\"u}rgen and S{\'e}n{\'e}gas, Julien}]{knopp2009iterative}
Knopp T, Eggers H, Dahnke H, Prestin J, S{\'e}n{\'e}gas J.
\newblock Iterative off-resonance and signal decay correction for improved
  multi-echo imaging in MRI.
\newblock IEEE Trans Med Imaging 2009;28(3):394--404.

\bibitem[{Pruessmann et~al.(1999)Pruessmann, Klaas P and Weiger, Markus and
  Scheidegger, Markus B and Boesiger, Peter}]{pruessmann1999sense}
Pruessmann KP, Weiger M, Scheidegger MB, Boesiger P.
\newblock SENSE: sensitivity encoding for fast MRI.
\newblock Magnetic resonance in medicine 1999;42(5):952--962.

\bibitem[{Uecker et~al.(2014)Uecker, Martin and Lai, Peng and Murphy, Mark J
  and Virtue, Patrick and Elad, Michael and Pauly, John M and Vasanawala,
  Shreyas S and Lustig, Michael}]{uecker2014espirit}
Uecker M, Lai P, Murphy MJ, Virtue P, Elad M, Pauly JM, et~al.
\newblock ESPIRiT—an eigenvalue approach to autocalibrating parallel MRI:
  where SENSE meets GRAPPA.
\newblock Magnetic resonance in medicine 2014;71(3):990--1001.

\bibitem[{Hestenes and Stiefel(1952)Hestenes, Magnus Rudolph and Stiefel,
  Eduard}]{hestenes1952methods}
Hestenes MR, Stiefel E.
\newblock Methods of conjugate gradients for solving linear systems.
\newblock Journal of Research of the National Bureau of Standards 1952;49(1).

\bibitem[{Beck and Teboulle(2009)Beck, Amir and Teboulle, Marc}]{beck2009fista}
Beck A, Teboulle M.
\newblock A fast iterative shrinkage-thresholding algorithm for linear inverse
  problems.
\newblock SIAM journal on imaging sciences 2009;2(1):183--202.

\bibitem[{Parikh et~al.(2014)Parikh, Neal and Boyd, Stephen and
  others}]{parikh2014proximal}
Parikh N, Boyd S, et~al.
\newblock Proximal algorithms.
\newblock Foundations and Trends{\textregistered} in Optimization
  2014;1(3):127--239.

\bibitem[{Goldstein and Osher(2009)Goldstein, Tom and Osher,
  Stanley}]{goldstein2009split}
Goldstein T, Osher S.
\newblock The split Bregman method for L1-regularized problems.
\newblock SIAM journal on imaging sciences 2009;2(2):323--343.

\bibitem[{Pruessmann et~al.(2001)Pruessmann, Klaas P and Weiger, Markus and
  B{\"o}rnert, Peter and Boesiger, Peter}]{pruessmann2001advances}
Pruessmann KP, Weiger M, B{\"o}rnert P, Boesiger P.
\newblock Advances in sensitivity encoding with arbitrary k-space trajectories.
\newblock Magnetic Resonance in Medicine: An Official Journal of the
  International Society for Magnetic Resonance in Medicine 2001;46(4):638--651.

\bibitem[{Knopp et~al.(2007)Knopp, Tobias and Kunis, Stefan and Potts,
  Daniel}]{knopp2007note}
Knopp T, Kunis S, Potts D.
\newblock A note on the iterative MRI reconstruction from nonuniform k-space
  data.
\newblock International journal of biomedical imaging 2007;2007.

\bibitem[{Pipe and Menon(1999)Pipe, James G and Menon,
  Padmanabhan}]{pipe1999sampling}
Pipe JG, Menon P.
\newblock Sampling density compensation in MRI: rationale and an iterative
  numerical solution.
\newblock Magnetic Resonance in Medicine: An Official Journal of the
  International Society for Magnetic Resonance in Medicine 1999;41(1):179--186.

\bibitem[{Inati et~al.(2017)Inati, Souheil J and Naegele, Joseph D and Zwart,
  Nicholas R and Roopchansingh, Vinai and Lizak, Martin J and Hansen, David C
  and Liu, Chia-Ying and Atkinson, David and Kellman, Peter and Kozerke,
  Sebastian and others}]{inati2017ismrm}
Inati SJ, Naegele JD, Zwart NR, Roopchansingh V, Lizak MJ, Hansen DC, et~al.
\newblock ISMRM Raw data format: A proposed standard for MRI raw datasets.
\newblock Magnetic resonance in medicine 2017;77(1):411--421.

\bibitem[{Fessler and Sutton(2003)Fessler, Jeffrey A and Sutton, Bradley
  P}]{fessler2003nonuniform}
Fessler JA, Sutton BP.
\newblock Nonuniform fast Fourier transforms using min-max interpolation.
\newblock IEEE transactions on signal processing 2003;51(2):560--574.

\bibitem[{Knopp(2014)Knopp, Tobias}]{knopp2014experimental}
Knopp T.
\newblock Experimental multi-threading support for the Julia programming
  language.
\newblock In: Proceedings of the 1st First Workshop for High Performance
  Technical Computing in Dynamic Languages IEEE Press; 2014. p. 1--5.

\bibitem[{Dagum and Menon(1998)Dagum, Leonardo and Menon,
  Ramesh}]{dagum1998openmp}
Dagum L, Menon R.
\newblock OpenMP: an industry standard API for shared-memory programming.
\newblock IEEE computational science and engineering 1998;5(1):46--55.

\bibitem[{Blumofe et~al.(1996)Blumofe, Robert D and Joerg, Christopher F and
  Kuszmaul, Bradley C and Leiserson, Charles E and Randall, Keith H and Zhou,
  Yuli}]{blumofe1996cilk}
Blumofe RD, Joerg CF, Kuszmaul BC, Leiserson CE, Randall KH, Zhou Y.
\newblock Cilk: An efficient multithreaded runtime system.
\newblock Journal of parallel and distributed computing 1996;37(1):55--69.

\bibitem[{Kukanov and Voss(2007)Kukanov, Alexey and Voss, Michael
  J}]{kukanov2007foundations}
Kukanov A, Voss MJ.
\newblock The Foundations for Scalable Multi-core Software in Intel Threading
  Building Blocks.
\newblock Intel Technology Journal 2007;11(4).

\bibitem[{Frigo and Johnson(1998)Frigo, Matteo and Johnson, Steven
  G}]{frigo1998fftw}
Frigo M, Johnson SG.
\newblock FFTW: An adaptive software architecture for the FFT.
\newblock In: Proceedings of the 1998 IEEE International Conference on
  Acoustics, Speech and Signal Processing, ICASSP'98 (Cat. No. 98CH36181),
  vol.~3 IEEE; 1998. p. 1381--1384.

\bibitem[{Ong et~al.(2018)Ong, F and Amin, S and Vasanawala, S and Lustig,
  M}]{ong2018mridata}
Ong F, Amin S, Vasanawala S, Lustig M.
\newblock Mridata. org: An open archive for sharing MRI raw data.
\newblock In: Proc. Intl. Soc. Mag. Reson. Med, vol.~26; 2018. p.~1.

\bibitem[{Maier et~al.(2020)Maier, Oliver and Baete, Steven Hubert and Fyrdahl,
  Alexander and Hammernik, Kerstin and Harrevelt, Seb and Kasper, Lars and
  Karakuzu, Agah and Loecher, Michael and Patzig, Franz and Tian, Ye and
  others}]{maier2020cgsense}
Maier O, Baete SH, Fyrdahl A, Hammernik K, Harrevelt S, Kasper L, et~al.
\newblock CG-SENSE revisited: Results from the first ISMRM reproducibility
  challenge.
\newblock Magnetic Resonance in Medicine 2020;.

\bibitem[{Eggers et~al.(2002)Eggers, H and Boernert, P and Boesiger,
  P}]{eggers2002comparison}
Eggers H, Boernert P, Boesiger P.
\newblock Comparison of gridding-and convolution-based iterative reconstruction
  algorithms for sensitivity-encoded non-Cartesian acquisitions.
\newblock In: Proceedings of the 10th Annual Meeting of ISMRM, Honolulu; 2002.
  p. 743.

\bibitem[{Gr{\"a}ser et~al.(2019)Gr{\"a}ser, Matthias and Thieben, Florian and
  Szwargulski, Patryk and Werner, Franziska and Gdaniec, Nadine and Boberg,
  Marija and Griese, Florian and Hofmann, Martin and Ludewig, Peter and Ven,
  Dionys van de and others}]{graser2019human}
Gr{\"a}ser M, Thieben F, Szwargulski P, Werner F, Gdaniec N, Boberg M, et~al.
\newblock Human-sized magnetic particle imaging for brain applications.
\newblock Nature communications 2019;10:1--9.

\bibitem[{Block and Wiggins(????)Block, Kai T and Wiggins, Roy}]{Yarra}
Block KT, Wiggins R, Yarra - A Toolbox for Clinical MRI Research;.
\newblock Accessed: 2020-01-20.
\newblock \url{http://http://yarraframework.com}.

\bibitem[{Borisch et~al.(2019)Borisch, Eric A and Grimm, Roger C and Riederer,
  Stehpen J.}]{Autorec}
Borisch EA, Grimm RC, Riederer SJ.
\newblock Automated Reconstruction Processing.
\newblock In: Proc. Intl. Soc. Mag. Reson. Med, vol.~27; 2019. p.~1.

\end{thebibliography}
\end{document}